\documentclass[aps,prl,twocolumn,superscriptaddress,floatfix]{revtex4}
\usepackage{graphicx}
\usepackage{multirow}

\begin{document} 
\title{Search for the Rare Decays $K_{L}\rightarrow\pi^{0}\pi^{0}\mu^{+}\mu^{-}$ and $K_{L}\rightarrow\pi^{0}\pi^{0}X^{0}\rightarrow\pi^{0}\pi^{0}\mu^{+}\mu^{-}$}


\affiliation{
University of Arizona, Tucson, Arizona 85721}
\affiliation{
University of California at Los Angeles, Los Angeles, California 90095}
\affiliation{
Universidade Estadual de Campinas, Campinas, Brazil 13083-970}
\affiliation{
The Enrico Fermi Institute, The University of Chicago, Chicago, Illinois
60637}
\affiliation{
University of Colorado, Boulder Colorado 80309}
\affiliation{
Elmhurst College, Elmhurst, Illinois 60126}
\affiliation{
Fermi National Accelerator Laboratory, Batavia, Illinois 60510}
\affiliation{
Osaka University, Toyonaka, Osaka 560-0043 Japan}
\affiliation{
Rice University, Houston, Texas 77005}
\affiliation{
Universidade de Sao Paulo, Sao Paulo, Brazil 05315-970}
\affiliation{
University of Virginia, Charlottesville, Virginia 22904}
\affiliation{
University of Wisconsin, Madison, Wisconsin 53706}
\author{
E.~Abouzaid}
\affiliation{
The Enrico Fermi Institute, The University of Chicago, Chicago, Illinois
60637}
\author{
M.~Arenton}
\affiliation{
University of Virginia, Charlottesville, Virginia 22904}
\author{
A.R.~Barker}
\altaffiliation[Deceased.]{}
\affiliation{
University of Colorado, Boulder Colorado 80309}
\author{
L.~Bellantoni}
\affiliation{
Fermi National Accelerator Laboratory, Batavia, Illinois 60510}
\author{
E.~Blucher}
\affiliation{
The Enrico Fermi Institute, The University of Chicago, Chicago, Illinois
60637}
\author{
G.J.~Bock}
\affiliation{
Fermi National Accelerator Laboratory, Batavia, Illinois 60510}
\author{
E.~Cheu}
\affiliation{
University of Arizona, Tucson, Arizona 85721}
\author{
R.~Coleman}
\affiliation{
Fermi National Accelerator Laboratory, Batavia, Illinois 60510}
\author{
M.D.~Corcoran}
\affiliation{
Rice University, Houston, Texas 77005}
\author{
B.~Cox}
\affiliation{
University of Virginia, Charlottesville, Virginia 22904}
\author{
A.R.~Erwin}
\affiliation{
University of Wisconsin, Madison, Wisconsin 53706}
\author{
C.O.~Escobar}
\affiliation{
Universidade Estadual de Campinas, Campinas, Brazil 13083-970}
\author{
A.~Glazov}
\affiliation{
The Enrico Fermi Institute, The University of Chicago, Chicago, Illinois
60637}
\author{
A.~Golossanov}
\affiliation{
University of Virginia, Charlottesville, Virginia 22904}
\affiliation{
Fermi National Accelerator Laboratory, Batavia, Illinois 60510}
\author{
R.A.~Gomes}
\altaffiliation[Permanent address: ]{IF-UFG - Goias, Brazil.}
\affiliation{
Universidade Estadual de Campinas, Campinas, Brazil 13083-970}
\author{
P.~Gouffon}
\affiliation{
Universidade de Sao Paulo, Sao Paulo, Brazil 05315-970}
\author{
Y.B.~Hsiung}
\affiliation{
Fermi National Accelerator Laboratory, Batavia, Illinois 60510}
\author{
D.A.~Jensen}
\affiliation{
Fermi National Accelerator Laboratory, Batavia, Illinois 60510}
\author{
R.~Kessler}
\affiliation{
The Enrico Fermi Institute, The University of Chicago, Chicago, Illinois
60637}
\author{
K.~Kotera}
\affiliation{
Osaka University, Toyonaka, Osaka 560-0043 Japan}
\author{
A.~Ledovskoy}
\affiliation{
University of Virginia, Charlottesville, Virginia 22904}
\author{
P.L.~McBride}
\affiliation{
Fermi National Accelerator Laboratory, Batavia, Illinois 60510}
\author{
E.~Monnier}
\altaffiliation[Permanent address: ]{C.P.P. Marseille/C.N.R.S., France.}
\affiliation{
The Enrico Fermi Institute, The University of Chicago, Chicago, Illinois
60637}
\author{
H.~Nguyen}
\affiliation{
Fermi National Accelerator Laboratory, Batavia, Illinois 60510}
\author{
R.~Niclasen}
\affiliation{
University of Colorado, Boulder Colorado 80309}
\author{
D.G.~Phillips~II}
\email[Correspondence should be addressed to David Graham Phillips II at ]{dgp@email.unc.edu}
\affiliation{
University of Virginia, Charlottesville, Virginia 22904}
\author{
H.~Ping}
\affiliation{
University of Wisconsin, Madison, Wisconsin 53706}
\author{
E.J.~Ramberg}
\affiliation{
Fermi National Accelerator Laboratory, Batavia, Illinois 60510}
\author{
R.E.~Ray}
\affiliation{
Fermi National Accelerator Laboratory, Batavia, Illinois 60510}
\author{
M.~Ronquest}
\affiliation{
University of Virginia, Charlottesville, Virginia 22904}
\author{
E.~Santos}
\affiliation{
Universidade de Sao Paulo, Sao Paulo, Brazil 05315-970}
\author{
W.~Slater}
\affiliation{
University of California at Los Angeles, Los Angeles, California 90095}
\author{
D.~Smith}
\affiliation{
University of Virginia, Charlottesville, Virginia 22904}
\author{
N.~Solomey}
\affiliation{
The Enrico Fermi Institute, The University of Chicago, Chicago, Illinois
60637}
\author{
E.C.~Swallow}
\affiliation{
The Enrico Fermi Institute, The University of Chicago, Chicago, Illinois
60637}
\affiliation{
Elmhurst College, Elmhurst, Illinois 60126}
\author{
P.A.~Toale}
\affiliation{
University of Colorado, Boulder Colorado 80309}
\author{
R.~Tschirhart}
\affiliation{
Fermi National Accelerator Laboratory, Batavia, Illinois 60510}
\author{
C.~Velissaris}
\affiliation{
University of Wisconsin, Madison, Wisconsin 53706}
\author{
Y.W.~Wah}
\affiliation{
The Enrico Fermi Institute, The University of Chicago, Chicago, Illinois
60637}
\author{
J.~Wang}
\affiliation{
University of Arizona, Tucson, Arizona 85721}
\author{
H.B.~White}
\affiliation{
Fermi National Accelerator Laboratory, Batavia, Illinois 60510}
\author{
J.~Whitmore}
\affiliation{
Fermi National Accelerator Laboratory, Batavia, Illinois 60510}
\author{
M.J.~Wilking}
\affiliation{
University of Colorado, Boulder Colorado 80309}
\author{
R.~Winston}
\affiliation{
The Enrico Fermi Institute, The University of Chicago, Chicago, Illinois
60637}
\author{
E.T.~Worcester}
\affiliation{
The Enrico Fermi Institute, The University of Chicago, Chicago, Illinois
60637}
\author{
M.~Worcester}
\affiliation{
The Enrico Fermi Institute, The University of Chicago, Chicago, Illinois
60637}
\author{
T.~Yamanaka}
\affiliation{
Osaka University, Toyonaka, Osaka 560-0043 Japan}
\author{
E.D.~Zimmerman}
\affiliation{
University of Colorado, Boulder Colorado 80309}
\author{
R.F.~Zukanovich}
\affiliation{
Universidade de Sao Paulo, Sao Paulo, Brazil 05315-970}

\collaboration{The KTeV Collaboration}
\noaffiliation

\begin{abstract} 
The KTeV E799 experiment has conducted a search for the rare
decays $K_{L}\rightarrow\pi^{0}\pi^{0}\mu^{+}\mu^{-}$ and $K_{L}\rightarrow\pi^{0}\pi^{0}X^{0}\rightarrow\pi^{0}\pi^{0}\mu^{+}\mu^{-}$, where the $X^{0}$ is a possible new neutral boson that was reported by the HyperCP experiment with a mass of (214.3$\pm$0.5) MeV/{\it c}$^2$.  
We find no evidence for either decay.  We obtain upper limits of Br($K_{L}\rightarrow\pi^{0}\pi^{0}X^{0}\rightarrow\pi^{0}\pi^{0}\mu^{+}\mu^{-}) < 1.0 \times 10^{-10}$ and Br($K_{L}\rightarrow\pi^{0}\pi^{0}\mu^{+}\mu^{-}) < 9.2 \times 10^{-11}$ at the 90\% confidence level.  This result rules out the pseudoscalar $X^{0}$ as an explanation of the HyperCP result under the scenario that the $\bar{d}$s$X^{0}$ coupling is completely real.  
\\ \\
\vspace{0.25in}
\noindent
PACS numbers: 13.20.Eb, 13.25.Es
\end{abstract}

\maketitle

The HyperCP collaboration has reported the possible observation of an $X^0$ boson of mass (214.3$\pm$0.5) MeV/{\it c}$^2$ decaying into $\mu^+\mu^-$ based on three observed events in a search for the decay $\Sigma^{+}\rightarrow p\mu^{+}\mu^{-}$ [1].  The confidence level within the Standard Model for all three events to overlap within the HyperCP dimuon mass resolution of 0.5 MeV/{\it c}$^2$ is less than 1\%.  As the $X^0$ would presumably be a strange-to-down neutral current, it is natural to look for it in the kaon sector, specifically in the mode $K_{L}\rightarrow\pi^{0}\pi^{0} X^{0}\rightarrow\pi^{0}\pi^{0}\mu^{+}\mu^{-}$.  This letter presents the first attempt to detect the rare decay modes $K_{L}\rightarrow\pi^{0}\pi^{0}\mu^{+}\mu^{-}$ and $K_{L}\rightarrow\pi^{0}\pi^{0}X^{0}\rightarrow\pi^{0}\pi^{0}\mu^{+}\mu^{-}$.  

Using a two-quark flavor changing coupling model in which the $X^{0}$ couples to $\bar{d}$s and $\mu^{+}\mu^{-}$, theoretical estimates of the $K_{L}\rightarrow\pi^{0}\pi^{0}X^{0}\rightarrow\pi^{0}\pi^{0}\mu^{+}\mu^{-}$ branching ratio were determined for a pseudoscalar $X^{0}$ and an axial vector $X^{0}$ [2].  Reference [2] uses the known value Br($K^{\pm}\rightarrow\pi^{\pm}\mu^{+}\mu^{-}$) = 8.1$\times$10$^{-8}$ [3] to rule out the possibility of a scalar or vector $X^{0}$ as explanations of the HyperCP anomaly.  These predictions assume real $\bar{d}$s$X^{0}$ couplings, ${\it g_{P}}$; for a complex coupling with a dominant imaginary term, $|\Im (g_{P})| > 0.98|g_{P}|$, the predicted upper limit is much smaller [4].  Another prediction of Br($K_{L}\rightarrow\pi^{0}\pi^{0}X^{0}\rightarrow\pi^{0}\pi^{0}\mu^{+}\mu^{-}$) for a pseudoscalar $X^{0}$ has been made [5].  Finally, the branching ratio for $K_{L}\rightarrow\pi^{0}\pi^{0}X^{0}\rightarrow\pi^{0}\pi^{0}\gamma\gamma$ has been estimated using a sgoldstino model [6].  These results are summarized in Table I.  

The E391a collaboration has reported [7] an upper limit Br($K_{L}\rightarrow\pi^{0}\pi^{0}X^{0}\rightarrow\pi^{0}\pi^{0}\gamma\gamma) < 2.4\times 10^{-7}$, which rules out the sgoldstino model of this decay.  The possibility [8] that $X^{0}$ could be a light pseudoscalar Higgs boson of the next-to-minimal supersymmetric Standard Model (NMSSM) was investigated at ${\it e}^{+}{\it e}^{-}$ colliders by CLEO [9] and BaBar [10,11,12] and at the Tevatron (D0) [13].  No evidence for a NMSSM light pseudoscalar Higgs boson was found.  

The $K_{L}\rightarrow\pi\pi X^{0}$ modes have an extremely limited phase space.  The phase space of $K_{L}\rightarrow\pi^{0}\pi^{0}X^{0}$ is ten times larger than the phase space available to $K_{L}\rightarrow\pi^{+}\pi^{-}X^{0}$, motivating the search for the former over the latter.  We have searched for $K_{L}\rightarrow\pi^{0}\pi^{0}\mu^{+}\mu^{-}$ and $K_{L}\rightarrow\pi^{0}\pi^{0}X^{0}\rightarrow\pi^{0}\pi^{0}\mu^{+}\mu^{-}$ in data from the 1997 and 1999 runs of KTeV E799 II at Fermi National
Accelerator Laboratory. 

The KTeV E799 experiment produced neutral kaons via collisions of 800 GeV/{\it c} protons with
a BeO target.  The particles created from interactions with the target passed through a series of collimators, absorbers and sweeper magnets to produce two nearly parallel $K_{L}$ beams.  The $K_{L}$ beams then entered a 65 m long vacuum tank, which was evacuated to 1 $\mu$Torr.  A doubling of the spill length and an increase in instantaneous luminosity between the 1997 and 1999 data-taking periods resulted in a factor of 2-3 increase in per-spill protons on target in 1999. A 10.3 inch beryllium absorber was introduced in the 1999 data-taking period to reduce neutron backgrounds. 

Immediately downstream of the vacuum region was a spectrometer composed of an analysis magnet between two pairs of drift chambers. The momentum kick imparted by the magnetic field was reduced from the 1997 value of 0.205 GeV/{\it c} to 0.150 GeV/{\it c} in 1999 to increase the acceptance for low 
momentum charged particles. The momentum resolution of the spectrometer in 1997 was $\sigma_{\it P}/{\it P} = 0.38\%\oplus 0.016\%{\it P}$ [14], and in 1999 the momentum resolution was $\sigma_{\it P}/{\it P} = 0.52\%\oplus 0.022\%{\it P}$.

\begin{table}
\begin{center}
\begin{tabular}{|c|c|}
  \hline
$X^{0}\rightarrow\mu^{+}\mu^{-}$ Model & Br($K_{L}\rightarrow\pi^{0}\pi^{0}X^{0}$) \\ \hline
Pseudoscalar ($\Re (g_{P})$) [2] & (8.3$^{+7.5}_{-6.6}$)$\times$10$^{-9}$ \\ \hline
Axial Vector ($\Re (g_{A})$) [2] & (1.0$^{+0.9}_{-0.8}$)$\times$10$^{-10}$ \\ \hline
Pseudoscalar ($|\Im (g_{P})| > 0.98|g_{P}|$) [4] & $< 7\times 10^{-11}$ \\ \hline
Pseudoscalar ($\Re (g_{P})$) [5] & 8.02$\times$10$^{-9}$ \\ \hline
sgoldstino ($X^{0}\rightarrow\gamma\gamma$) [6] & 1.2$\times 10^{-4}$ \\ \hline
\end{tabular}
\caption{Summary of predicted branching ratios for $K_{L}\rightarrow\pi^{0}\pi^{0}X^{0}$.}
\end{center}
\end{table}

The electromagnetic calorimeter was constructed of 3100 pure CsI crystal blocks arranged into a 1.9$\times$1.9 m$^{2}$ array.  Each CsI crystal was 27 radiation lengths long.  Two holes were located near the center of the calorimeter to allow for passage of the beams.  The electromagnetic calorimeter had an energy resolution of $\sigma_{E}/E \simeq 0.4\% \oplus 2\% /\sqrt{E[GeV]}$ and the position resolution was about 1 mm.  The muon ID system used a Pb wall, three steel filters and three scintillator counter planes to identify muons by filtering out other charged particles.  The muon ID system contained 31 nuclear interaction lengths of material and had a charged pion fake rate of (1.69 + 0.17{\it P} [GeV/{\it c}])$\times 10^{-3}$, where {\it P} is the track momentum.  A photon veto system detected photons outside the detector acceptance.  The upstream section of the photon veto system had five lead-scintillator counter arrays located inside the vacuum decay region.  The downstream section of the photon vetos had four lead-scintillator arrays that framed the outside of the last three drift chambers and the CsI calorimeter.  A more detailed description of the KTeV detector and photon veto system can be found in [15, 16, 17].        

The signal modes and normalization mode ($K_L\rightarrow\pi^0\pi^0\pi^0_{D}$, where one photon was lost down the beam hole and $\pi^0_D\rightarrow e^{+}e^{-}\gamma$) were collected by different triggers.  The triggers required in-time energy clusters in the calorimeter of at least 1 GeV.  The signal mode required one (two) such clusters for the 1997 (1999) data-taking periods.  Two hits were required in the two most downstream stations of the muon system in 1997; in 1999, the number of hits required in the middle station was reduced to one.  The normalization mode trigger required at least four in-time clusters and two tracks.  

Both tracks were required to form a good vertex within the vacuum decay region, to match (within 7 cm) a cluster in the CsI calorimeter and to deposit less than 1 GeV of energy in the CsI calorimeter, consistent with a muon hypothesis.  99.9$\%$ of muons with a track momentum over 7.0 GeV/{\it c} satisfied the last three requirements.  Each of the three scintillator counting planes in the muon ID system were required to register at least one hit.  The invariant $\mu^{+}\mu^{-}$ mass, $M_{\mu\mu}$, was required to be less than 0.232 GeV/{\it c}$^{2}$, which is slightly above the kinematic limit given by $M_{K}$ - 2$M_{\pi}$.            

Four clusters in the calorimeter without associated tracks were required.  The resolution of the z-vertex determined from the two $\gamma\gamma$ vertices associated with a $\pi^0 \pi^0$ was better than the resolution of the z-vertex from the two muons.  We considered each possible $\gamma\gamma$ pair to find the combination with the best agreement between the positions of the two $\gamma\gamma$ decay points under the hypothesis that each originated from a $\pi^0$ decay.  A minimum pairing chi-squared, $\chi_{z}^{2}$, was calculated to determine the best agreement between the positions of the two $\gamma\gamma$ decay points.  A weighted average of z-vertex values for each $\gamma\gamma$ in the pairing with the minimum  $\chi_{z}^{2}$ was used as the decay vertex for the event.  This vertex was then required to be located within the length of the vacuum decay region.  A $\gamma\gamma$ mass, $M_{\gamma\gamma}$, was calculated for the event using the decay vertex from the minimum $\chi_{z}^{2}$ pairing.  $M_{\gamma\gamma}$ was required to be within 0.009 GeV/{\it c}$^2$ of the $\pi^0$ mass. 

The $K_{L}\rightarrow\pi^{0}\pi^{0}\mu^{+}\mu^{-}$ simulation was modeled as a four body decay using a constant matrix element.   The $K_{L}\rightarrow\pi^{0}\pi^{0}X^{0}\rightarrow\pi^{0}\pi^{0}\mu^{+}\mu^{-}$ simulation was modeled as a three body decay with a flat phase space, where the $X^{0}$ underwent a prompt decay to $\mu^{+}\mu^{-}$.
\begin{figure}[hbtp]
  \begin{center}
    \scalebox{1.0}{\includegraphics[height=5.13cm,width=8.35cm]{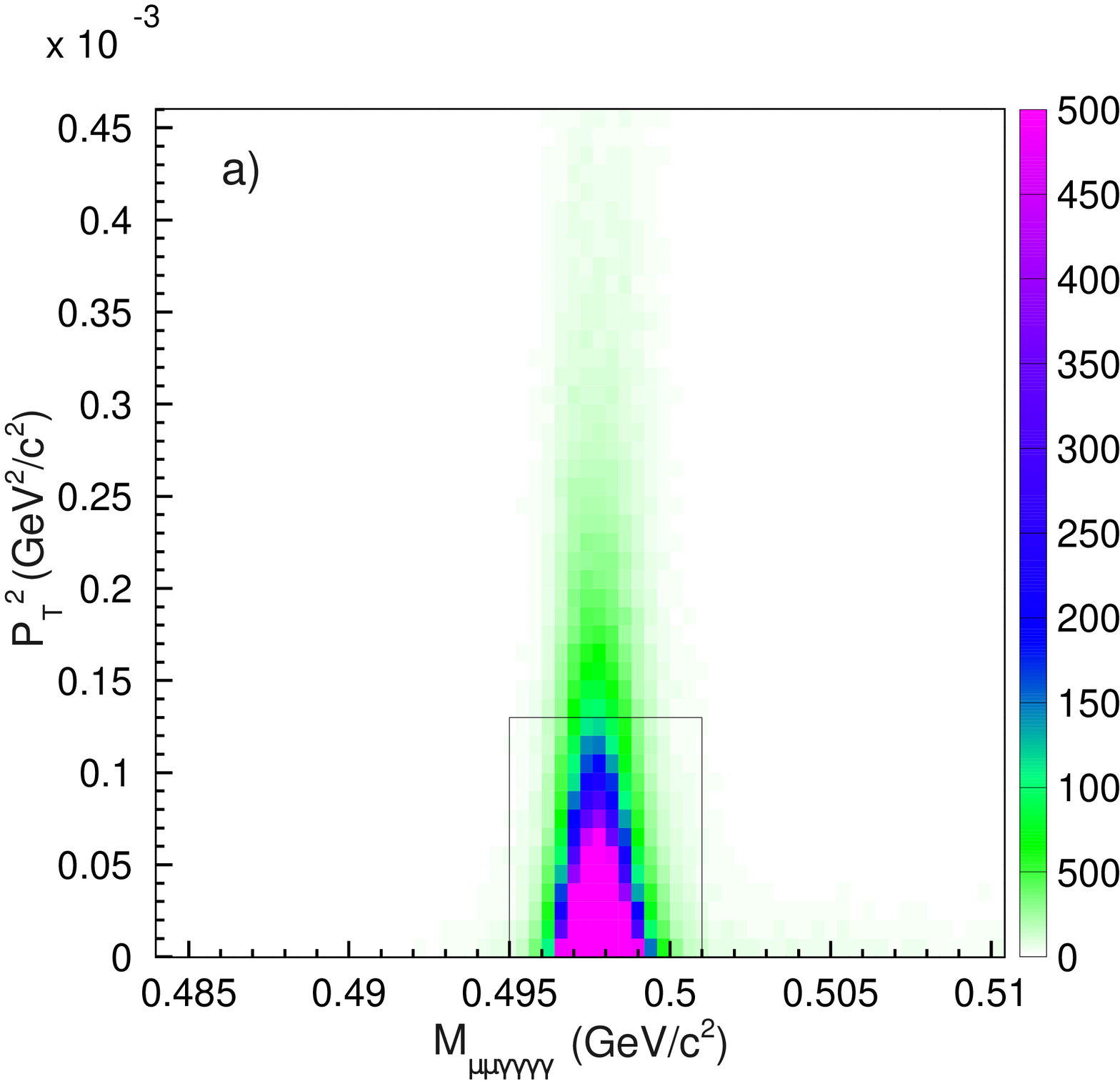}}
    \scalebox{1.0}{\includegraphics[height=5.13cm,width=8.35cm]{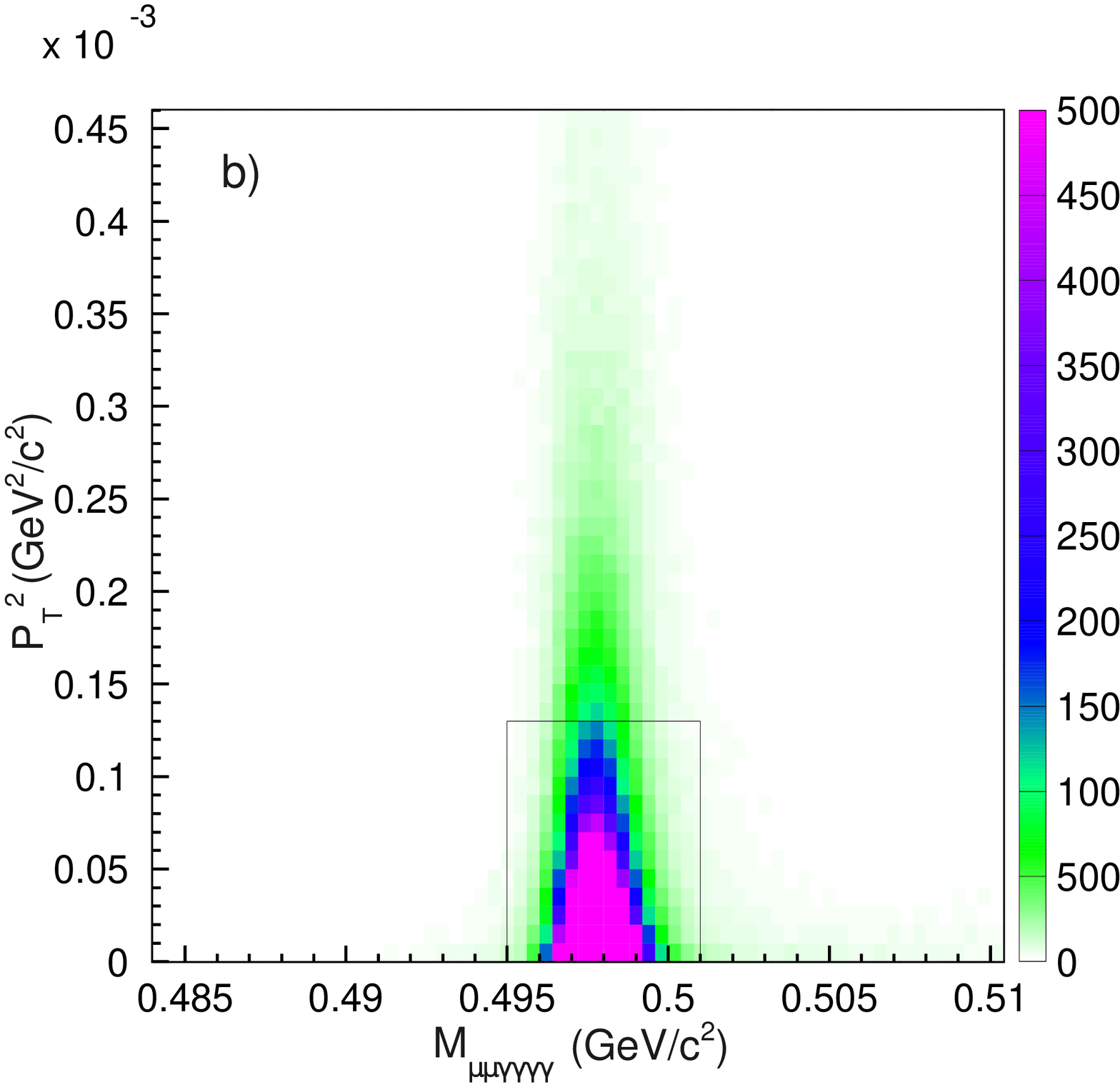}}
    \scalebox{1.0}{\includegraphics[height=5.13cm,width=8.35cm]{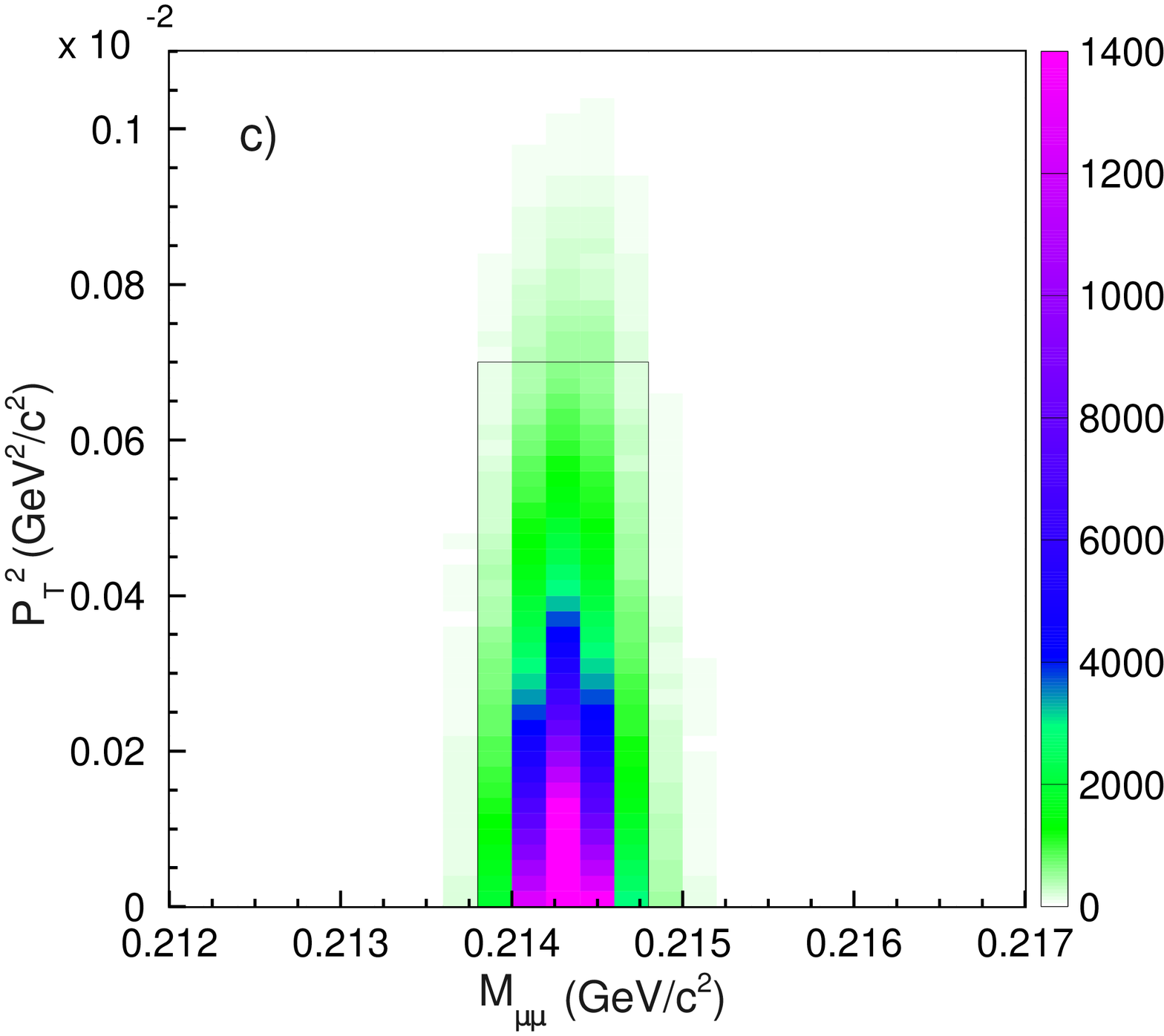}}
    \scalebox{1.0}{\includegraphics[height=5.13cm,width=8.35cm]{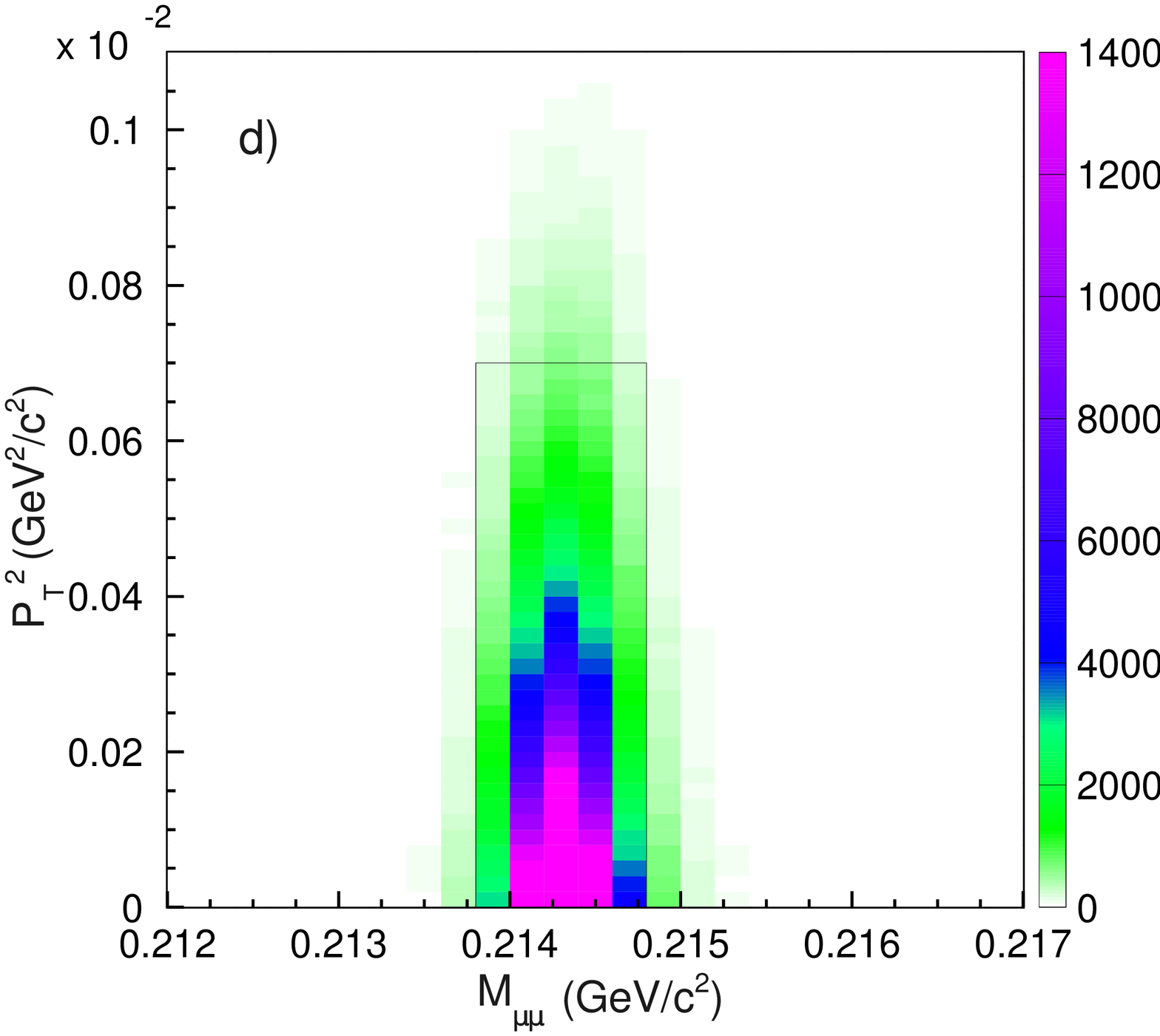}}
    \caption{a,b)  $p_{T}^{2}$ vs. $M_{\mu\mu\gamma\gamma\gamma\gamma}$ for the 1997 and 1999 $K_{L}\rightarrow\pi^{0}\pi^{0}\mu^{+}\mu^{-}$ simulation respectively.  The boxed signal region contains 90$\%$ of all events.  c,d)  $|p^2_T(\mu\mu) - p^2_T(\gamma\gamma\gamma\gamma)|$ vs. $M_{\mu\mu}$ for the 1997 and 1999 $K_{L}\rightarrow\pi^{0}\pi^{0}X^{0}\rightarrow\pi^{0}\pi^{0}\mu^{+}\mu^{-}$ simulation respectively.  The boxed signal region contains 95$\%$ of all events.  The four plots are shown after all analysis requirements were applied.}
  \end{center}
\end{figure}

The signal regions for the 1997 and 1999 data were based on the $M_{\mu\mu\gamma\gamma\gamma\gamma}$, $p^2_T(\mu\mu\gamma\gamma\gamma\gamma)$ and $|p^2_T(\mu\mu) - p^2_T(\gamma\gamma\gamma\gamma)|$ resolutions calculated in the simulation.  Here $p^2_T$ is
measured transverse to the direction of the $K_L$, determined by the line
connecting the BeO target and the vertex.  For a well-measured decay, $p^2_T(\mu\mu\gamma\gamma\gamma\gamma)$ and $|p^2_T(\mu\mu) - p^2_T(\gamma\gamma\gamma\gamma)|$ should be close to zero.  The $K_{L}$ signal region for the decay $K_{L}\rightarrow\pi^{0}\pi^{0}\mu^{+}\mu^{-}$ was defined as 0.495 GeV/{\it c}$^2\leq M_{\mu\mu\gamma\gamma\gamma\gamma}
\leq$ 0.501 GeV/{\it c}$^2$ and $p_T^2\leq$ 1.3$\times 10^{-4}$ (GeV/{\it c})$^2$.  The $X^{0}$ signal region for the $K_{L}\rightarrow\pi^{0}\pi^{0}X^{0}\rightarrow\pi^{0}\pi^{0}\mu^{+}\mu^{-}$ decay was defined as 213.8$\times 10^{-3}$ GeV/c$^2\leq M_{\mu\mu}\leq$ 214.8$\times 10^{-3}$ GeV/{\it c}$^2$ and $|p^2_T(\mu\mu) - p^2_T(\gamma\gamma\gamma\gamma)|\leq$ 7.0$\times 10^{-4}$ (GeV/{\it c})$^2$.  The bound on $M_{\mu\mu}$ was determined from the conservative hypothesis that the observations made by HyperCP reflect the natural width of the $X^{0}$ [18].  Figure 1 shows $p_T^2$ vs. invariant mass plots from the $K_{L}\rightarrow\pi^{0}\pi^{0}\mu^{+}\mu^{-}$ and $K_{L}\rightarrow\pi^{0}\pi^{0}X^{0}\rightarrow\pi^{0}\pi^{0}\mu^{+}\mu^{-}$ signal mode simulations.         

Every $K_L$ decay mode with two minimum ionizing tracks and at least one photon was considered as a potential source of background.  Accidental time-coincident activity created from particle interactions in the vacuum window, neutrons from the target, cosmic rays, beam interactions or another kaon decay in flight can overlap with the primary kaon decay in an event to reproduce the signal mode topology.  We have simulated all known backgrounds to the extent possible.  Accidental activity was included in the simulation of all background mode events.  Small branching ratio backgrounds such as $K_{L}\rightarrow\pi^{0}\pi^{\pm}\mu^{\mp}\nu_{\mu}$ [19] and $K_{L}\rightarrow\pi^{+}\pi^{-}\gamma$ were simulated with the full statistics of the data.  Large branching ratio modes such as $K_{L}\rightarrow\pi^{\pm}\mu^{\mp}\nu_{\mu}$ were also studied extensively, although simulated samples with statistics similar to the data were not feasible.  We find that when accidental activity reproduces the signal mode topology, $p_T^2$ and the invariant mass $M_{\mu\mu\gamma\gamma\gamma\gamma}$ are pushed to values well above the signal region.  The conclusion that the background is negligible is confirmed in the data.

The normalization mode shares the topological trait of four photons and two tracks with the signal mode and has a well-understood branching ratio.  The vertex in the normalization mode analysis was required to be located within the vacuum decay region.  The signal region for the 1997 data was defined by 0.494 GeV/{\it c}$^2$ $\leq$ $M_{ee\gamma\gamma\gamma\gamma}$ $\leq$ 0.501 GeV/{\it c}$^2$ and $p_T^2 (ee\gamma\gamma\gamma\gamma)$ $\leq$ 0.00015 GeV$^2$/{\it c}$^2$.  The signal region for the 1999 data was a contour that was derived from a joint probability distribution of $M_{ee\gamma\gamma\gamma\gamma}$ and $p_T^2 (ee\gamma\gamma\gamma\gamma)$ signal resolutions from simulations.  

The flux, $F_{K}$, obtained from the normalization mode, is the estimated number of $K_L$ decays in the vacuum decay region.  Uncertainties in $F_{K}$ originated from the branching ratio used to calculate $F_{K}$ and the muon ID system efficiency.  The uncertainty due to normalization mode requirements was studied by varying the selection requirements in simulation and data and noting the change in the estimated flux. The uncertainty due to differences in the signal and normalization mode simulations was estimated by varying the selection requirements of both modes in the simulation.  The statistical uncertainties on the signal mode simulation were less than 0.14$\%$ for each decay mode.  The statistical uncertainty for the normalization mode simulation was less than 0.37$\%$, while the statistical uncertainty for the normalization mode data was less than 1.14$\%$.  Systematic uncertainty in the muon ID efficiency came from modeling of the energy loss in the muon filters and from simulation of gaps between scintillator paddles in the muon planes [20].  A systematic uncertainty associated with the muon trigger inefficiency was determined by selecting clean $K_{\mu 3}$ decays from a minimum bias trigger [21].  Results from these systematic uncertainty studies are given in Table II. 
\begin{table}
\begin{center}
\begin{tabular}{|c|c|}
  \hline
Systematic Uncertainty on $F_{K}$ & $\frac{\Delta F_{K}}{F_{K}}$ \\ \hline 
\multirow{2}{*}{Variation of Normalization} & \\
Requirements & 3.57$\%$ \\ \hline
\multirow{2}{*}{Variation of Signal/Normalization} & \\
Requirements & 5.35$\%$ \\ \hline
Muon Trigger Inefficiency & 2.00$\%$ \\ \hline
Cracks in Muon Counting Planes & 0.50$\%$ \\ \hline
Energy Loss in Muon Filters & 0.40$\%$ \\ \hline
$Br(K_{L}\rightarrow\pi^{0}\pi^{0}\pi^{0})$ [3] & 0.61$\%$ \\ \hline
$Br(\pi^{0}\rightarrow\gamma\gamma)$ [3] & 0.03$\%$ \\ \hline
$Br(\pi^{0}\rightarrow e^{+}e^{-}\gamma)$ [3] & 2.98$\%$ \\ \hline
Total Systematic Uncertainty & 7.41$\%$ \\ \hline
\end{tabular}
\caption{Summary of systematic uncertainties on the apparent $K_L$ flux, labeled as $F_{K}$.}
\end{center}
\end{table}

\begin{figure}[hbtp]
  \begin{center}
    \scalebox{0.44}{\includegraphics{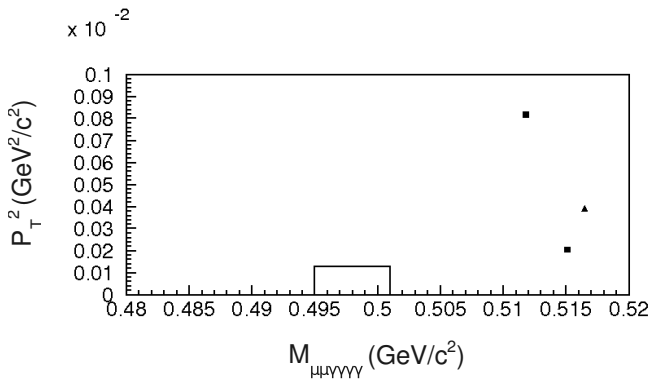}}
    \caption{$p_{T}^{2}$ vs. $M_{\mu\mu\gamma\gamma\gamma\gamma}$ plot for the combined 1997 and 1999 data sets, which are indicated by triangles and squares respectively.  The signal box is open.}
  \end{center}
\end{figure}

The 1997 (1999) signal mode acceptance was 3.14$\%$ (4.03$\%$) and 2.80$\%$ (3.74$\%$) for $K_{L}\rightarrow\pi^{0}\pi^{0}\mu^{+}\mu^{-}$ and $K_{L}\rightarrow\pi^{0}\pi^{0}X^{0}\rightarrow\pi^{0}\pi^{0}\mu^{+}\mu^{-}$ respectively.  The 1997 and 1999 normalization mode acceptances were 4.21$\times 10^{-6}$ and 3.26$\times 10^{-6}$ respectively [22].  While signal mode acceptance would drop in scenarios for which the $X^{0}$ does not decay immediately, it would not be sharply reduced for values of $c\tau <$ 3 mm, where $\tau$ is the proper decay time.  $F_{K}$ was 2.73$\times 10^{11}$ in 1997 and 4.12$\times 10^{11}$ in 1999.  The single event sensitivity was 3.97$\times 10^{-11}$ for $K_{L}\rightarrow\pi^{0}\pi^{0}\mu^{+}\mu^{-}$ and 4.34$\times 10^{-11}$ for $K_{L}\rightarrow\pi^{0}\pi^{0}X^{0}\rightarrow\pi^{0}\pi^{0}\mu^{+}\mu^{-}$.  Figure 2 displays the results of the blind analysis; no events are inside the signal regions after opening the signal boxes and no events were found within the available $\mu^{+}\mu^{-}$ phase space.  Using the method of [23], the 90$\%$ confidence level upper limits are $Br(K_{L}\rightarrow\pi^{0}\pi^{0}\mu^{+}\mu^{-}) <  9.2 \times 10^{-11}$ and $Br(K_{L}\rightarrow\pi^{0}\pi^{0}X^{0}\rightarrow\pi^{0}\pi^{0}\mu^{+}\mu^{-}) < 1.0 \times 10^{-10}$.

Our result for $Br(K_{L}\rightarrow\pi^{0}\pi^{0}X^{0}\rightarrow\pi^{0}\pi^{0}\mu^{+}\mu^{-})$ is nearly two orders of magnitude smaller than the expected branching ratios for $K_{L}\rightarrow\pi^{0}\pi^{0}X^{0}\rightarrow\pi^{0}\pi^{0}\mu^{+}\mu^{-}$ from [2] and [5], in which $X^{0}$ was taken to be a pseudoscalar.  This rules out the pseudoscalar $X^{0}$ as an explanation of the HyperCP result under the premise that ${\it g_{P}}$ is completely real and also places a tight bound on ${\it g_{P}}$ of $|\Im (g_{P})| \gtrsim 0.98|g_{P}|$ [4].  Finally, our upper limit challenges the axial-vector $X^{0}$ explanation of the HyperCP result.

We thank the Fermi National Accelerator Laboratory staff for their contributions. This work was supported by the U.S. Department of Energy, the U.S. National Science Foundation, the Ministry of Education and Science of Japan, the Funda\c c\~ao de Amparo a
Pesquisa do Estado de Sao Paulo-FAPESP, the Conselho Nacional de
Desenvolvimento Cientifico e Tecnologico-CNPq, and the CAPES-Ministerio da Educa\c c\~ao.

\end{document}